\documentclass{pinchcr}   

\usepackage{graphicx}        
\usepackage{amsmath}

\renewcommand{\phi}{\varphi}

\newcommand{\be}{\begin{equation}}
\newcommand{\ee}{\end{equation}}

\title{Scientific interview}

\author{Jorge Kurchan}
\affiliation{PMMH, ESPCI, 10 rue Vauquelin, CNRS UMR 7636 , Paris, France 75005}

\author{James S. Langer}
\affiliation{Physics Department, University of California,
Santa Barbara, CA 93106, USA
}

\author{Thomas A. Witten}
\affiliation{The James Franck Institute and The Department of Physics,
The University of Chicago, 929 East 57th Street, Chicago, Illinois 60637}

\author{Peter G. Wolynes}
\affiliation{Department of Chemistry and Biochemistry, University of California at San Diego, La Jolla, California 92093-0371, USA}

\begin{document}

\maketitle

\preface

Four leading scientists in the field of the glass transition answer
to a series of questions formulated by the Editors (L. Berthier, G.
Biroli, J.-P. Bouchaud, L. Cipelletti, W. van Saarloos). No specific
format has been imposed to their answers: the scientists were free
to answer or not to any given question, using how much space they
felt appropriate. After a first round of answers, each author was
given the opportunity to read the answers by his colleagues and to
adjust his own answers accordingly.

The questions were:
\begin{itemize}

\item Q1) In your view, what are the most important aspects of the experimental data on the glass transition that any consistent theory explain? Is dynamical
heterogeneity one of these core aspects?

\item Q2) Why should we expect anything universal in the behaviour of glass-forming liquids? Is the glass transition problem well defined? Or is any glassy liquid glassy
in its own way?

\item Q3) In spin glasses, the existence of a true spin glass phase transition has been well established by simulations and experiments. Do you believe that a similar result will ever be demonstrated for molecular glasses?

\item
Q4) Why are there so many different theories of glasses? What kind of decisive
experiments you suggest to rule out at least some of them?

\item
Q5) Can you briefly explain, and justify, why you believe your pet theory fares better than others? What, deep inside, are you worried about, that could jeopardize your theoretical construction?

\item
Q6) In the hypothesis that RFOT forms a correct skeleton of the theory of glasses, what is missing in the theoretical construction that would convince the community?

\item
Q7) Exactly solvable mean-field glass models exhibit an extraordinary complexity
requiring impressive mathematical tools to solve them. Are you worried that for both the spin and structural glass problems it proves so hard to establish the
validity of mean-field concepts in finite dimensions? Why are finite dimensional versions of mean-field glasses always behaving in a very non mean-field manner?

\item
Q8) In your view, do the recent ideas and experimental developments concerning jamming in
granular media and colloids contribute to our understanding of molecular
glasses, or are they essentially complementary?

\item
Q9) If a young physicist asked you whether he or she should work on the glass problem in the next few years,
would you encourage him or her and if so, which aspect
of the glass problem would you recommend him or her to tackle? If no, what problem in condensed matter theory should
he or she tackle instead? If yes, what
particular aspect of the glass problem seems the most exciting at present?

\item
Q10) In twenty years from now, what concepts, ideas or results obtained on the glass transition in the last twenty years will be remembered?

\item
Q11) If you met an omniscient God and were allowed one single question on glasses, what would it be?

\end{itemize}

\section{Jorge Kurchan answers}

Note: only references to papers that are
difficult to identify are given here.  For a general presentation, see the review
article \shortcite{cavagna09}.

\vspace{.7cm}

{\em
Q1) In your view, what are the most important aspects of the experimental data on the glass transition that any consistent theory explain? Is dynamical
heterogeneity one of these core aspects?
}

\vspace{.7cm}
 The mere existence of heterogeneities is not in itself an additional fact,  but a mathematical necessity: there have to be
heterogeneities if there is  a timescale that grows faster than Arrhenius in a system
of soft particles.
The only way a system  can construct  zero modes or higher barriers is through cooperativity. Of course, if one
has access to the detailed space-time form of the heterogeneities, then this is extra information.

A model of fragile glasses should capture  the relation between growing timescales and vanishing  entropy.
I would add that fluctuation-dissipation relations should reproduce the Lennard-Jones results of Barrat and Berthier~\cite{Barrat_Berthier},
a useful discriminant since the  precise form of these relations is model-dependent. All these aspects are subject to the uncertainties
associated with short times.

If we make a multi-dimensional scatter plot of the  parameters describing existing glass formers, we shall find that there are correlations.
Those between fragility, specific heat  and stretching exponents have been discussed  by Wolynes and coworkers, and there must
be others I do not know. To the extent that they are significant, correlations must be explained. If they are imperfect, this has to be
explained too.

Having said this,  I think that all in all it will be the internal  logic and the tractability  of the theory -- {\em the fact that the
objects it invokes and relates  have a clear definition} -- that will make it satisfactory.

\vspace{.7cm}

{\em
Q2) Why should we expect anything universal in the behaviour of glass-forming liquids? Is the glass transition problem well defined? Or is any glassy liquid glassy
in its own way?
}

\vspace{.7cm}

Here one should separate two levels. The word ``glassy'' is used  (a) in the general sense of slow (or {\em slowing}) dynamics.  (b) for ``fragile'' glasses with their rather specific features of entropy and timescales.

The general sense (a) includes systems such as the two and three dimensional spin glasses (including those having Potts or $p$-spin interactions) ,
``dirty'' ferromagnets,  crystal ripening and defect annealing, quasicrystal annealing,  strong and fragile glasses, and also kinetically constrained
models, frustration limited domains, ``Backgammon'',  ``car-parking'' , Tower of Hanoi, and a myriad  other models.
The class of fragile glasses (b) is much more specific: it includes a  group of systems having  slowing down of dynamics in a characteristic (non-power law) super-Arrhenius way, with what seems to be a concomitant decrease in entropy.  Polymer and some molecular glasses
are of this kind.

In the general class (a) there might be many ways of being glassy. One might still expect  some universal features  that are just a consequence of the dynamics being   slow, or entropy production small.  There are relations between timescale increase and dynamics correlations, bounds for the violations of the fluctuation-dissipation relation, generic features of response of the system to driving, etc, that apply to every slow system. We would  like to know much more, as this might be a front where out of equilibrium thermodynamics might develop -- it being less ambitious than the  generic nonequilibrium problem.

Class (b) is much more restrictive, to the point that some  consider it empty.
Super-Arrhenius behavior -- the logarithm of the timescale  growing faster than linearly in the inverse temperature --
implies growing dynamic {\em and static} lengthscales (see Q1) , and this leads to the search of some form of hidden order,  with perhaps
complete ordering eventually avoided.   Here the expectation is that things are much more universal in the sense of applying to
 less systems but for those,  essentially in the same  way. Nobody will
be happy with a  mechanism, an order parameter or a lengthscale,  that is defined differently for polymer and a monoatomic  glasses.

\vspace{.7cm}

{\em
Q3) In spin glasses, the existence of a true spin glass phase transition has been well established by simulations and experiments. Do you believe that a similar result will ever be demonstrated for molecular glasses?
}

\vspace{.7cm}
The question  whether there is a spin-glass transition  in a magnetic field is still very much unresolved, and is the object of an impassioned and contradictory  literature.
 The spin-glass example is interesting, as it illustrates a new development that we are witnessing: simulation timescales are for the first time overlapping experimental ones. What in fact we discover is that there is quite a good continuity between  experiments and simulations, {\em and that there are issues that neither will resolve. }

 To illustrate how simulations may have a paradoxical effect,
 consider the example of the JANUS aging simulations ~\cite{JANUS}. They measure two-time correlation functions of a $3D$ Edwards-Anderson model,
 up to times corresponding  $0.1$ seconds of experiments. They propose a form for the long-time correlations:
 \begin{equation}
 C(t,t_w)= A(t_w) \; \left(1+\frac{t}{t_w} \right)^{-1/\alpha(t_w)}
 \end{equation}
 The factor $A(t_w)$ is familiar from experimental fits: it is important -- and unknown -- whether it goes to zero or not as  $t_w \rightarrow \infty$.
  If $A(t_w)$  goes to zero with $t_w$ it means that the Edwards-Anderson parameter is zero  (in a ferromagnet
$A(t_w) \rightarrow 0$   would be interpreted as evidence for  Kosterlitz-Thouless like order).  If this were the case for spin-glasses,
  {\em neither} the droplet {\em nor}
the replica symmetry breaking scenario would apply in three dimensions, so it is quite an issue. Up to now it was thought that the time-dependence of  $A(t_w)$ in experiments might  be a consequence of the spins being not quite Ising-like,  but the JANUS  work have proven that similar results are observed in simulations, where the nature of the spins is obviously under control.
These simulations cannot  resolve conclusively the value of $\lim_{t_w \rightarrow \infty}  A(t_w)$.
  This might sound like a paradox, but what they have done, in 'validating' the experimental results, is to prove  that even experimental times are too short  to resolve the issue.

  In fact, it has been argued on the basis of extrapolating simulation results that the longest correlation lengths achieved in spin glass experiments are about   twenty 'lattice sites' (see Fig. 1 of \cite{BB}).   In conclusion: simulations have convinced us that experimental times are short.

(Somebody  may  ask:  why should we care, then, about longer times and lengthscales? Is it legitimate to take more seriously  a length  because we think
  it would  ideally continue to grow in times beyond the  experimental ones?)

\vspace{.7cm}

{\em
Q4) Why are there so many different theories of glasses? What kind of decisive
experiments you suggest to rule out at least some of them?
}

\vspace{.7cm}

 As I mentioned above, in many systems (e.g. colloids, spin glasses) simulations  have quite literally caught up with experiments.
Note how different the situation is  in the field of strongly interacting electrons, where nature thermalizes easily
what computers cannot simulate --  a consequence  of the fact that nature often
solves the 'sign problem' much better than computers, while it is as bad as computers in implementing  glassy dynamics.

A new, decisive experiment, will probably be a  though experiment, somewhat like question Q11. The one we have suggested with D. Levine is
to consider a large snapshot of a glass configuration, as equilibrated as possible.
Take patches of size $\ell$ and look for places where they repeat to some significant  precision. Is there a  length $\ell_o$
such that there is a crossover from
a high frequency of repetition for patch sizes  $\ell< \ell_o$ to an exponentially  low, essentially random frequency of repetition for $\ell>\ell_o$?
Does $\ell_o$ grow as the glass ages? Is it infinite in an equilibrium  ideal glass? (as it is in all
other forms of order we know: crystalline, quasicrystalline, amorphous tilings...)
Note that this definition of the length $\ell_o$ is model-independent, cfr. the end of Q2.

\vspace{.7cm}

{\em
Q5) and Q6) Can you briefly explain, and justify, why you believe your pet theory fares better than others? What, deep inside, are you worried about, that could jeopardize your theoretical construction? In the hypothesis that RFOT forms a correct skeleton of the theory of glasses, what is missing in the theoretical construction that would convince the community?
}

\vspace{.7cm}

In the glass literature there are plenty of phenomenological arguments,
 of useful representations, of metaphors - sometimes accompanied by a model
that illustrates them.
 RFOT is unique in that it has some microscopic grounds, it is quite constrained (we cannot add or remove pieces at will), and, above
all, has time and again given results that were not expected. Consider the following (the 'you' here is a rhetoric device, it refers to no-one in particular):

\begin{itemize}

\item  You read Derrida's paper on the Random Energy Model (REM) and you  realize that
the transition is precisely the one  proposed  by Kauzmann  for an ideal glass (at $T_K$), with the random energy levels playing the role
of states.

\item  You are unsatisfied with the REM because it is somewhat ad hoc. So, recognizing that the REM is an example of the {\em one step replica symmetry breaking} kind, you look in the literature for other more microscopic models of this kind.
You find the  mean-field Potts glass, spin glasses with multi-spin interactions, etc.

\item
 These models being microscopic, you may as well study their dynamics. Remarkably, in the high temperature phase, the dynamics turns out to
be described by
the Mode-Coupling (MCT) equations, which are currently studied for glasses. There is a Mode Coupling Transition at a certain $T_d>T_K$,  known already to be an artifact of the approximation. Within your perspective you understand perfectly well why the MCT transition has to go away in finite
dimensions, and only one at the  Kauzmann transition may remain - thus forcing you to identify the point at which viscosity diverges
 $T_o$ with the thermodynamic temperature $T_K$.

\item You are however uncomfortable with having used models with quenched disorder, so you try mean-field 'glass' models without quenched disorder.
You find that almost every model you can construct has a static and dynamic behavior just like the ones you have studied,  something you can prove within mean-field.
A new surprise is that  often an isolated deep
state appears that is given by number theoretical properties (see~\cite{MPR1} for a number-theoretic solution and~\cite{MPR2} for the mean-field version of the fully-frustrated model):   you have unexpectedly encountered what looks like a mean-field version of the crystalline state.

\item
 Now that you have a microscopic model, you are not forced to restrict the dynamics
  to the high temperature phase. So you quench the models to low temperatures, and you find that the system does not equilibrate: it 'ages', just
  like true glasses. When you look at the properties of observables, you discover that the slow fluctuations behave as if they were  'thermalized' in an effective
  temperature of the order of the one of the glass transition. You discover that 'fictive temperatures'  have been around since the 1940's, and it is likely
  that what you have discovered is precisely that -- minus the exhilarating freedom of phenomenological discourse. In particular, you
  see explicitly that the effective temperatures are {\em not} just a consequence of the system staying in a
  pre-quench  configuration: {\em 'hot' backbones are reconstructed time and again during aging}.
You also study how the system responds to forces that do work on it:  you find the generic phenomenology of 'shear thinning' of supercooled liquids,
 and in some cases
you can explain the much more rare 'shear thickening' ~\cite{Mauro} of certain glasses.

\item Now you go back to the energy and free energy landscape, something that you could not do in a pure mode coupling context.
  You find that the main features (importance and location of saddles, marginality, etc)
   are extremely close to those discussed many years ago by Goldstein, to the point that you wonder whether he secretly  knew the solution you have.

\end{itemize}

 So far, you are encouraged by what seems an amazing unification, such as happen in other, more fortunate fields of physics. In fact, you begin to think
 that
a substantial part of the  twentieth century phenomenological thinking was a non-mathematical construction of a mean-field picture. This
  also implies that the  mean field picture  inherits
the criticisms to these ideas.
At this point, however, you begin to encounter difficulties.
You know that the dynamic (MCT) transition you have obtained is  a mean-field illusion, one that has preempted (and thus hidden from you)
  the ideal glass one. Furthermore, stable metastable states   of free energy density higher than the fundamental (as you invoked to
  define configurational entropy) are somewhat arbitrarily defined beyond
  mean-field.
  You conclude that an important part  of the theory of glassy dynamics will be, unfortunately, {\em even
qualitatively} beyond mean-field.

\begin{itemize}

\item
  Technically speaking, this means that you have to calculate corrections to
the dynamics that are {\em non-perturbative} in the mean-field parameter: these are the ``activated processes''.
A true calculation of these is beyond everyone's capabilities, but an estimate for a {\em finite size} mean-field system
gives you a timescale that diverges at $T_K$ in a way
that does not  quite conform
to a Vogel-Fulcher law. You need to move   more resolutely into finite dimensions.

\item More trouble: when you attempt to simulate finite dimensional versions of the very models that inspired  you (random Potts, p-spin)
nothing like a Kauzmann transition appears.
Worse: some models without quenched disorder that worked well at the mean field level
(e.g. the fully frustrated~\cite{MPR1,MPR2} model) becomes quite different from a glass in finite dimensions.

Your problem is not that you do not have three-dimensional models of fragile glasses  without quenched disorder
(there are plenty, hard spheres for example), but that these
do not interpolate easily with a mean-field version which you can control (See last question in Q7).
An example that works, studied by Dotsenko  ~\cite{Dotsenko} is that of particles in $d$ dimensions interacting with an oscillatory (cosine) potential, with, say,
a finite range cutoff. As the range tends to infinity, one obtains exactly a case of mean-field RFOT.

\item
If you are interested in thermodynamics (as opposed to dynamics) you may still use a mean-field approach with spatial degrees of freedom,
somewhat in the same spirit of the local mean-field approximations for ferromagnetic systems. The numbers obtained for the glass transition
and the configurational entropy are quite encouraging. The whole picture for the thermodynamics of hard sphere glasses is also
quite satisfactory (see, however, the caveat in  Q7(b)).
In the same spirit, you can work with variable-range interactions, with the long-range limit being
the mean-field theory.

\item
But for true dynamics with 'activated processes' you need to plunge into true finite dimensions. Because you cannot do anything analytically, you work phenomenologically, with the mean-field theory in mind. (Your
phenomenologist friends, whom you so criticized before,  are gloating).
Above the Kauzmann temperature the  metastable (now unstable) states of mean-field break into pieces, the 'mosaics'.
One invokes a ``library'' of mosaics (with, it turns out, extremely brief ``books'' -- the size of a Haiku) that can occupy a given volume, and argues on the basis of an interplay between the entropy associated to their multiplicity and their free energy.
You obtain this way a form for the relaxation time, and suggestions of relations between parameters (cfr. Q1) that
turn out to be  rather well respected, and, much more
importantly, sometimes unexpected.

\item
One last worry. For temperatures at which the  dynamics has virtually ground to a halt, you interpret this as a consequence
of  a particle cooperation that extends to  a few  (six at best, three in the computer)  inter-particle distances. It seems that many models
will be able to achieve this,  without a  true coherence length that is building up --  and how is one to distinguish? Are you not
defending the mosaics on the basis of what they would become, if given unrealistically large times?

\end{itemize}

This should not be the end of the story. The mosaic phenomenology may be considered as a promising start, but a lot  will depend on to what extent finite-dimensional
RFOT can rise above phenomenology in practice.
RFOT is a remarkable building, with the  upper floor made of wood for lack of budget.  The fact that the top floor  stands out as a different, temporary
construction  is just a proof
of the ambitious character  of the whole enterprize.

\vspace{.7cm}

{\em
Q7a) Exactly solvable mean-field glass models exhibit an extraordinary complexity
requiring impressive mathematical tools to solve them. Are you worried that for both the spin and structural glass problems it proves so hard to establish the
validity of mean-field concepts in finite dimensions? }

\vspace{.7cm}

This is worrying. Mean field theory for fully connected systems is quite hard, both statically and dynamically. Extended  to randomly dilute
(Viana-Bray like) systems, it becomes daunting. Even analytic results need considerable numerical effort to extract numbers from them.
What will  become of this in finite dimensions? Who is going to do this, and if somebody can, who  will be able to read it?
This is all the more worrying because the force of RFOT  in mean-field version is that it its pieces are
quantitatively defined, so it is disappointing if in finite dimensions it loses this edge.

\vspace{.7cm}

{\em Q7b)Why are finite dimensional versions of mean-field glasses always behaving in a very non mean-field manner?}

\vspace{.7cm}

It would seem that finite dimensional models either keep you dynamically away from their Kauzmann
transition  (if you want to approach it in equilibrium), or they allow you to get close and then you see that it is not as you expected it: either there is no transition or it is spin-glass like (there are many articles on this, by S. Franz, Binder and co-workers, Katzgraber and  Young, on variants of disordered  $p$-spin and Potts models).
Does this mean that the Kauzmann transition is avoided in finite dimensions, mirroring the situation of the MCT dynamic transition?
In other words, is it possible that nonperturbative corrections to the  equilibrium calculation destroy the Kauzmann transition?
If this is so, we do not know through which  mechanism this would happen.  (Years ago, Stillinger proposed one, but it relies on
the identification of ``state'' with```inherent structure''.)

\vspace{.7cm}

{\em
Q8) In your view, do the recent ideas and experimental developments concerning jamming in
granular media and colloids contribute to our understanding of molecular
glasses, or are they essentially complementary?
}

\vspace{.7cm}

Systems of hard, frictionless particles in mechanical equilibrium are hypo or  isostatic,
the static equations are either  underdetermined or just determined - like a table with two or three
legs touching the floor. From this, one argues that they are extremely sensitive to perturbations, the  breaking of a particle contact
has a far-reaching effect.

 The sensitivity mentioned above
is manifested in the appearance of many soft vibration modes (which become manifest if the pressure is slightly released or the potential is softened),
 and divergent lengthscales in the hard/jammed limit.
 In conclusion,
the infinite pressure point for a system
of {\em hard} spheres is  a critical point.
The question is: is this critical point, and its divergent lengths, responsible for the ``glass order'', or is it something that {\em happens}
to a system that is  amorphous for other reasons?

The example of a  FCC crystal of spheres with very small polydispersity is eloquent. As soon as polydispersity is turned on, there is a proliferation of soft modes beyond the  usual acoustic modes of the crystal. At infinite pressure we have a critical point just as in an amorphous system. And yet, the nature of order which makes  the system  solid  is that of a crystal,
as soon as the pressure is decreased slightly.
This suggests that the same is true for amorphous systems: the nature of the order that makes
the system a solid  at finite pressures is
not related to the ``jamming'' criticality as found at infinite pressure. This is not to say that this criticality may not have important consequences  for a system
at finite pressures, and that it may be especially relevant for granular and even colloidal matter.

\vspace{.7cm}

{\em
Q9) If a young physicist asked you whether he or she should work on the glass problem in the next few years,
would you encourage him or her and if so, which aspect
of the glass problem would you recommend him or her to tackle? If no, what problem in condensed matter theory should
he or she tackle instead? If yes, what
particular aspect of the glass problem seems the most exciting at present?
}

\vspace{.7cm}

``If the nature of the present impasse you fully  understand,
and a contribution you still feel you can make, then go for it you must''

\vspace{.7cm}

{\em
Q10) In twenty years from now, what concepts, ideas or results obtained on the glass transition in the last twenty years will be remembered?
}

\vspace{.7cm}

It is difficult to say, as in the glass field an analytical success  often  leads to a relapse into even more vigorous phenomenological discourse,
the case of effective temperatures is a case in point.
There is a reason for this: one does not accept easily that something as familiar as glasses, which are not even quantal, should require
complicated mathematics.

 Is it possible that there will be new developments starting from the theory of low-dimensional amorphous order, as in non-periodic tilings?
It is amazing how little attention the glass community has paid to lessons that could be learned from quasicrystals and in general non-periodic systems. Perhaps rightly so, I do not know.

In any case, my feeling is that in twenty years time many perceived contradictions between different  pictures may
be seen as only apparent.
If you say {\em ``the glass transition is only dynamic''}, this might stir controversy. Instead: {\em ``I understand that a slowing timescale requires a growing  equilibrium correlation length, we know that in real systems
the dynamics has slowed down
considerably by the time this correlation length is three, so I wish to understand how this comes about at such modest equilibrium correlation lengths''} -- will be accepted by everyone.
Similarly, I would expect  that Mode Coupling theory will be eventually accepted as part of  RFOT, something that has not  yet happened,  perhaps for sociological reasons.

\vspace{.7cm}

{\em
Q11) If you met an omniscient God and were allowed one single question on glasses, what would it be?
}

\vspace{.7cm}

Are there general, non-trivial  principles in nonequilibrium thermodynamics (perhaps for the restricted case of slow dynamics)?
What I have in mind is something  like Prigogine's  Minimum Dissipation Principle, or Jaynes' Maximal Entropy Prescription -- {\em but true}.

\vspace{1cm}


\section{James S. Langer answers}

{\em
Q1) In your view, what are the most important aspects of the experimental data on the glass transition that any consistent theory explain? Is dynamical
heterogeneity one of these core aspects?
}

\vspace{.7cm}

I am not sure that dynamical heterogeneity is an absolutely essential aspect of glassy systems; but I hope it is, because it seems to emerge naturally from the physical pictures that make most sense to me.  I think the most important aspects that we need to understand are the dramatic slowing of molecular rearrangements that occurs with decreasing temperature, and the relation between this dynamic behavior and the thermodynamic behavior seen at about the same temperatures.

The difficulty in drawing definite conclusions about heterogeneity - from either experiments or numerical simulations - is that too many different mechanisms can be occurring and competing with each other at the same time.  These different phenomena are essentially impossible to disentangle because everything is happening so slowly in the most interesting circumstances.   For example, we saw numerical evidence at the meeting in Leiden that crystallites form during the cooling of some (all?) binary glass-forming liquids.  Does the system fail to complete a first order phase transition because the driving force for coarsening is too small at high temperatures?  Or because the kinetics becomes anomalously slow at lower temperatures?  Or are we seeing a slowly fluctuating equilibrium phase of crystallites in coexistence with eutectic fluids?  If so, is this a glass?

Another example - Harrowell~\shortcite{100,101} and coworkers have provided evidence that regions in which molecules are most likely to be mobile may be regions in which the underlying solidlike noncrystalline material is elastically soft, i.e. where there are localized low-frequency vibrational modes.  Is this ``dynamic heterogeneity?''  Or is it just an intrinsic property of any disordered material?

\vspace{.7cm}

{\em
Q2) Why should we expect anything universal in the behaviour of glass-forming liquids? Is the glass transition problem well defined? Or is any glassy liquid glassy
in its own way?
}

\vspace{.7cm}

I think we should keep open minds about these questions.  Our job as theorists is to look for general concepts, and we can't let ourselves be completely discouraged by the possibility that our ``universalities'' may not always be so universal, or that one supposedly ``universal'' behavior turns into another one as the temperature or other parameters are changed.

We've found some promising candidates for general concepts in recent decades.  For example, there is the ``frustration'' picture of a glass-forming liquid,  in which the energetically favored local correlations are inconsistent with ordered, space-filling states.  This picture implies some measure of universality in the intrinsic slowing-down process, because the frustration length scales ought to become irrelevant in comparison with some dynamic length scale (what dynamic length scale?) as the temperature decreases.   But this picture can be spoiled by competing equilibration processes such as growth of crystallites, or by a crossover from one dominant relaxation mechanism to another as a function of temperature.  The ways in which universality can be broken are likely to be strongly system dependent and, therefore, inconveniently complex.

Another potentially universal concept that I like increasingly well these days is the idea that the degree of disorder in a glassy material can be characterized by an effective temperature.  Jorge Kurchan has been one of the pioneers in this area.  (For example, see~\shortcite{200}. Some of my own recent work on this topic is described in \shortcite{300,400}.)  This idea is turning out to be especially useful for understanding nonequilibrium phenomena such as shear-banding instabilities and even possibly shear fracture.  A temperature-like measure of disorder that plays a central role in glass dynamics (sometimes equivalent to the role played by free volume) might be a definitively glassy feature, although it might also be useful for describing other disordered systems such as polycrystalline solids with high densities of dislocations and other defects.

\vspace{.7cm}

{\em
Q3) In spin glasses, the existence of a true spin glass phase transition has been well established by simulations and experiments. Do you believe that a similar result will ever be demonstrated for molecular glasses?
}

\vspace{.7cm}
I suspect that spin glasses, even those with short-ranged interactions, belong to a different universality class than molecular glasses.  In the latter case, I take seriously the simple argument that there can be no true glass transition.  In an indefinitely large, amorphous system with slightly soft interactions, there is always some way for two molecules to move around each other.  All that is needed is a thermally activated fluctuation in which the neighboring molecules squeeze out of the way and thereby open enough room for the molecular rearrangement to occur.  This process, or some energetically more efficient version of it,  may require an increasingly large amount of activation energy as the temperature decreases, because we must insist that the rearrangement results in a stable new configuration that doesn't just jump back to the original one during the next thermal fluctuation.  Thus, the spontaneous rearrangement rate may rapidly become unobservably slow at low temperatures, giving the appearance of a sharp transition to infinite relaxation time.  But mathematically there is no transition unless some kind of long-range order intervenes.  I don't think that that argument applies to spin glasses where spins occupy fixed positions and the random interactions are frozen.

In accord with my remarks in (Q2), however, I don't think that this argument - that there is no glass transition -  should stop us from trying to make theories of glasses. In fact, I think that this argument may be too academic to be useful. We know that glassy slowing down sets in and becomes increasingly pronounced at characteristic ``glass temperatures,'' which must correspond to characteristic energies and therefore characteristic dynamic mechanisms.  The proper goal of a theory is to identify these mechanisms and their ranges of validity.  The likelihood that they become irrelevant, or are dominated by other mechanisms at higher and lower temperatures, doesn't make it any less important to understand them

\vspace{.7cm}

{\em
Q4) Why are there so many different theories of glasses? What kind of decisive
experiments you suggest to rule out at least some of them?
}

\vspace{.7cm}

I'm not sure that there are really ``so many'' theories of glasses, at least not so many that try to be comprehensive, and that are sharply enough formulated to make unambiguous predictions.   In fact, I'm not sure that there are \emph{any} such theories.

There are, indeed, a great many \emph{models} of glassy behavior.  Most of these focus only on some but not all of the features that we think are characteristics of glasses.  By doing this, they have contributed a great deal to the conceptual tool kit that we use for interpreting data.  The kinetically constrained models are certainly in this category.  They have proven to be very useful phenomenologically; but by definition they focus primarily on kinetics as opposed to dynamics, and they haven't yet made a clear enough connection to molecular motions to satisfy me.  The mean-field models are the opposite.  They produce fascinating insights about thermodynamics and some kinds of glass-like transitions.  But, for reasons that I will say more about, I worry that they may be intrinsically incapable of addressing the most important dynamic questions. There are also various special-purpose models, designed to describe specific phenomena.  Here, I am thinking of the trapping models, or the anomalous diffusion model that several of us have used to interpret scattering experiments as well as diffusion and viscosity data.  These are all complementary and mostly non-contradictory approaches toward understanding glassy behavior.  Few of them pretend to be comprehensive theories.

There is also, of course, the wonderful world of glassy models that we can study by numerical simulation these days - hard spheres, soft spheres, molecules with all kinds of structures and local coordinations, even polymers.  Some of these simulations have played the roles of well controlled, benchmark experiments. Simulations become essential parts of the theoretical picture when they are carried out in direct connection with the development of theoretical models.

I am optimistic about the prospects for sorting out these different ideas by means of the increasingly powerful experimental tools that we have available these days.  My recommended projects, i.e. those that appeal to me as a theorist,  are the ones that bring a wide range of instrumental techniques to bear on a few well characterized glass-forming systems, and which go as close as possible to where we think there might be a glass transition.  As I said in Leiden, I'd like to see neutron scattering data for systems at temperatures close to the glass temperature, so that we can compare the scattering measurements with the data for viscosity and diffusion that are available there.
\vspace{.7cm}

{\em
Q5) Can you briefly explain, and justify, why you believe your pet theory fares better than others? What, deep inside, are you worried about, that could jeopardize your theoretical construction?
}

\vspace{.7cm}

The closest I come to having a ``pet theory'' is what I've called the ``excitation-chain mechanism''~\shortcite{600,601,700}.  I've tried to insist that this is not yet really a ``theory;'' it is at most a ``framework'' for a theory.  It's my best attempt so far to implement my bias about where the crux of the problem lies.

In my opinion, the glass problem is primarily a dynamics problem that needs to be addressed at the molecular level.  We should look at a noncrystalline array of molecules interacting via short-ranged forces, and ask why, and under what circumstances, it becomes anomalously difficult for these molecules to undergo rearrangements.  More specifically, we should choose some class of rearrangements, and compute the rates at which they are induced spontaneously by thermal fluctuations.  The typical rearrangement that I have in mind occurs when a molecule moves out of an energetically favored position, forming the glassy equivalent of a fully dissociated vacancy-interstitial pair, leaving a lower density at one place and producing a higher density somewhere else.  This is an important class of fluctuations because it determines diffusion and viscosity coefficients, and also is relevant to nonlinear irreversible behavior.  I hope that, once I understand this process, other pieces of the puzzle - including glassy thermodynamics - will fall together more easily.

Accordingly, I have asked what thermally activated molecular motions might most efficiently produce such irreversible fluctuations.   My philosophical model for this was the Fisher-Huse~\shortcite{800,801} estimate of relaxation rates in spin glasses, which makes a specific guess about the intermediate states that enable a correlated cluster of spins to change from one stable microstate to another.  I was guided more specifically by numerical simulations of molecular glass dynamics by Glotzer and coworkers~\shortcite{900}, in which chain-like motions appeared during thermally activated transitions between inherent structures.

The excitation-chain picture has allowed me to use little more than a back-of-the-envelope calculation to derive a Vogel-Fulcher formula for the transition rates.  I also have been able, with not much more than dimensional analysis, to connect this dynamic result to thermodynamic behavior near the glass temperature.  Most importantly for my recent research, the excitation chains have given me a way to understand how the rate factor for chain-enabled molecular rearrangements fits into theories of plastic deformation and failure.  These results emerge naturally from the basic picture, with few \emph{ad hoc} assumptions; but a systematic derivation of them, starting from molecular models, has eluded me so far. That's why I call this just the framework for a theory.

Do I worry that this could be completely wrong?  Of course I do.  I have tried to find a more systematic derivation, using function-space methods to compute probabilities for chains of molecular displacements in random environments.  I learned enough from the function-space effort to be sure that my estimates break down if I try to push them too far into the  low temperature regime.  I hope to come back to this effort some day.  At the moment, I can only guess that this is qualitatively the kind of theory we'll  need if we are ever going to have a real ``theory of glasses,'' i.e. a theory that starts with first principles and makes predictions.

\vspace{.7cm}

{\em
Q6) In the hypothesis that RFOT forms a correct skeleton of the theory of glasses, what is missing in the theoretical construction that would convince the community?
}

\vspace{.7cm}

A large part of my motivation for thinking about excitation chains was my discomfort with the random first-order transition (RFOT) theory~\shortcite{1000,1001,1002}. In my last evening in Leiden, Peter Wolynes and I spent several hours talking about these issues, and I hoped that we had found a point of view that would reduce our differences of opinion to well defined technical issues. Maybe we made some progress toward that goal.  Here's how I see the present state of the discussion.

I have been dubious about the RFOT theory, partly because it didn't seem to address the issues at what I thought was the right level dynamically, and partly because it seemed to be internally inconsistent.  These difficulties were most apparent in the earlier statements of the theory, where Wolynes and colleagues postulated that the transition state leading from one glassy configuration to another is an entropically favored droplet.  They called this droplet an ``instanton,'' and clearly were thinking about droplet models of first-order phase transitions.  Such models are well understood in conventional nucleation problems, where the initial state is a thermodynamically metastable phase, and the transition state contains a coexisting, marginally stable droplet of the nucleating phase.  The transition takes place when that droplet fluctuates over its activation barrier and starts to grow, converting the metastable system to a thermodynamically stable one.

The problem with the nucleation version of the RFOT theory is that the initial state is not a thermodynamic phase at all, but is one of an extensive number of microstates; and yet the droplet free energy is computed by summing over all of the microstates as if one were computing the thermodynamic free energy of the glass itself - which, in principle, is what is being done.  This is a manifestly non-orthodox theoretical procedure.  It is not meaningful to talk about coexistence or first-order phase transitions between microstates on the one hand and states of thermodynamic equilibrium on the other. Moreover, the entropic contribution to the droplet free energy is computed by making the Gibbsian assumption that a statistically significant fraction of the microstates is sampled in times shorter than any other time of interest.  But the time of interest here is precisely that sampling rate - the rate at which the system moves from one microstate to another - and the problem is to understand why that rate is so small.  Thus there seems to be an internal contradiction.

Recent presentations of the RFOT theory resemble an interesting reinterpretation by Bouchaud and Biroli~\shortcite{1100}. Their idea is to understand the sum over microstates in the droplet free energy as a sum over system trajectories leading from the initial state to all possible  final microstates within the spatial region occupied by the droplet. Each of these trajectories has a small \emph{a priori} probability of occurring.  But, if the region is big enough, there are enough of these trajectories that the probability of one of them being realized becomes appreciable. Accordingly, the transition state is the smallest region of a microstate that allows enough such trajectories that the region becomes unstable against a spontaneous molecular rearrangement.  The mathematics is unchanged from the original RFOT theory, but the language is quite different.

In rough outline, the Bouchaud-Biroli interpretation looks a little like droplet nucleation.  In fact, the interpretation of nucleation in terms of system trajectories is what I used in my ``instanton'' papers over forty years ago~\shortcite{1200,1201}. There, however, I knew that the individual molecular motions occur very rapidly compared to the rate at which a thermodynamically (almost) stable droplet changes its size and shape in response to thermal fluctuations. Therefore, I could use coarse-grained quantities such as surface energies and diffusion constants to model the motion of the droplet as it makes its way over the nucleation barrier.  In other words, I had a complete dynamic description of this transition state, including the relevant rate factors, and I knew that this description was appropriate for the time and length scales at which it was being used.

The reinterpreted RFOT theory has few such reliable ingredients.  No attempt is made to understand what fraction of the microstates within a region are actually accessible to the initial state.  The assumption is simply that they are all equally accessible and carry equal weight.  This assumption may be related to the surface-energy problem, which is the least well understood ingredient of the RFOT theory in any of its versions.  In order for the theory to make sense, the surface energy must scale like the square root of the number of molecules in the droplet instead of being proportional to the geometric surface area.  The arguments presented in support of this result are based on statistical analyses that I find hard to justify, especially in view of the fact that this transition state apparently contains no more than a few tens of molecules near the glass temperature, and considerably less at higher temperatures.  In contrast, the excitation-chain mechanism does not require that the transition state be a compact entity at all.  The square root of the number of molecules emerges naturally in the same place as in the RFOT theory, and the two analyses share many features beyond this point in their developments.

It must be said in its favor that the the RFOT theory identifies the right groups of parameters and puts them in the right places for comparison with experiment.  (The excitation-chain picture does this almost as well.)  My sense is that we are all on roughly the right track here, but that we are missing some theoretical tools.  Perhaps we just need a more powerful mathematical starting point.  A truly new idea might be better.

\vspace{.7cm}
{\em
Q7) Exactly solvable mean-field glass models exhibit an extraordinary complexity
requiring impressive mathematical tools to solve them. Are you worried that for both the spin and structural glass problems it proves so hard to establish the
validity of mean-field concepts in finite dimensions? Why are finite dimensional versions of mean-field glasses always behaving in a very non mean-field manner?}

\vspace{.7cm}

My worry about the mean-field models goes back to my experiences with the droplet calculations that I mentioned in (Q6).  I was using simple field theories; but it was clear that no perturbation expansion, no matter how many graphs it summed or how nonlinearly self-consistent it might be, could produce the interesting results, which always had essential singularities at the important places.  This is well known.  Such difficulties arise whenever the crucial physics involves statistically improbable, localized entities such as critical droplets, which don't naturally emerge in mean-field approximations, and especially not in the exactly solvable models.  The localized features are too easily lost when theories start with very long-ranged interactions or very high dimensionalities, and try to perturb away from those limits.

Therefore, I suspect that the exactly solvable mean-field theories are not going to come to grips with the dynamics of real molecular glasses near their glass temperatures.  Nor are the mode-coupling theories likely to be successful in this regard.  They do a good job of expanding away from liquidlike states, but break down at lower temperatures, just when the dynamics becomes most interesting.  It seems possible that we might figure out how to use such theories nonperturbatively and induce them to tell us about dynamic heterogeneities.  (Schweizer has made an important effort along these lines~\shortcite{1300}.)  Even then, however, I am afraid we would be starting at the wrong place.  If the transition state relevant to anomalously slow relaxation is, in fact, a delicately balanced, non-compact entity that involves only tens and not millions of molecules, then  conventional field theoretic approaches seem doomed to failure.

\vspace{.7cm}

{\em
Q8) In your view, do the recent ideas and experimental developments concerning jamming in
granular media and colloids contribute to our understanding of molecular
glasses, or are they essentially complementary?
}

\vspace{.7cm}

I think that the jamming ideas and the work on hard spheres and athermal granular systems have been enormously valuable, and am almost sure that there are deep connections between  these systems and molecular glasses.  For purposes of studying glass physics, we should exclude the complications of force chains and bridging configurations and the like that occur in static granular materials.  Even without these complications, however,  the connection between jamming transitions and glass transitions seems to be complex and subtle.  Understanding that connection could shed light on both kinds of phenomena.

\vspace{.7cm}

{\em
Q9) If a young physicist asked you whether he or she should work on the glass problem in the next few years,
would you encourage him or her and if so, which aspect
of the glass problem would you recommend him or her to tackle? If no, what problem in condensed matter theory should
he or she tackle instead? If yes, what
particular aspect of the glass problem seems the most exciting at present?
}

\vspace{.7cm}

The first part of this question - whether I encourage young scientists to go into any area of materials theory, especially in the United States these days - requires an essay by itself.  Let me say just that I am very grumpy about the state of funding in these areas, and even grumpier about the decline in fundamental research in areas such as materials engineering and materials chemistry.  Those fields once were homes for research in amorphous materials.  They are now seriously in need of new ideas, but basic research has mostly disappeared in them.

At the same time, however, I am enormously enthusiastic about the prospects for major progress in the nonequilibrium aspects of glass physics.  I started thinking about the glass problem only a few years ago, because I needed to understand some aspects of it in connection with my attempts to make theories of deformation and failure of noncrystalline materials.  Among other projects at the moment, I am working with two graduate students on applications of these theories to shear failure in earthquake faults.  We've found some fascinating results; and in this case the seismologists are gratifyingly receptive.

I admit that I am leery about encouraging young people to tackle the core problems in glass physics, where both the scientific and sociological barriers are uncomfortably high.  But there are great opportunities in related areas, especially including biological phenomena. In short, I think that this field has outstanding intellectual and practical potential, which is why I've chosen to work in it.

\vspace{.7cm}

{\em
Q10) In twenty years from now, what concepts, ideas or results obtained on the glass transition in the last twenty years will be remembered?
}

\vspace{.7cm}

Primarily, I hope to be surprised by the answer to this question. However, I find it hard to believe that we'll have forgotten some of the basic phenomenologies and concepts that have emerged in recent decades.  For example, I think that the Kauzmann paradox, the Adam-Gibbs connection between dynamics and some kind of entropy, the definition of inherent structures, the concept of fragility, and the like will have lasting value. But I hope that we'll have found new ways to think about these phenomena in another decade or so.

\vspace{.7cm}

{\em
Q11) If you met an omniscient God and were allowed one single question on glasses, what would it be?
}

\vspace{.7cm}

After having written my answers to some of these questions, especially (5) and (6), I think I'd ask this God to tell me what mathematical tools provide a natural language for talking about glassy dynamics problems.  I'm afraid, however, that His/Her reply would be something like this:

``\&*\%\$* you Langer!  I've already given you a computer.  Anyway, you should know by now that not every problem has a tractable mathematical solution.  You've already seen lots of examples of no-solution problems in nuclear structure physics, astrophysics, turbulence, etc. where the best you can do is compute special cases.''

Then I'd meekly ask, if I were allowed to do so - ``But haven't you made it seem as if there is something universal about glassy dynamics? And, if there is something universal, doesn't that mean that there is something fundamental to be discovered, and some natural way to describe it?''

To this, I'm afraid that She/He would answer:``Maybe''

\vspace{1cm}


\section{Thomas A. Witten answers}

{\em
Q1) In your view, what are the most important aspects of the experimental data on the glass transition that any consistent theory explain? Is dynamical
heterogeneity one of these core aspects?
}

\vspace{.7cm}
Form of divergent dynamic modulus with temperature or density.
Dynamical heterogeneity is one of these core aspects insofar as their
behavior has common features of many systems.  Molecular mobility
near walls or other constrained regions.

\vspace{.7cm}

{\em
Q2) Why should we expect anything universal in the behaviour of glass-forming liquids? Is the glass transition problem well defined? Or is any glassy liquid glassy
in its own way?
}

\vspace{.7cm}

This point of view says that the local motions that allow flow in
a system where motion is highly constrained are dictated by the local
molecular packing and bonding geometry.  This point of view might
well be sufficient to explain all practical measurements.  Likewise,
the world might have been such that we could never approach a
critical temperature more closely than ten percent.  In that case we
could never make an experimentally decisive demonstration of critical
universality.  For features of critical phenomena like the operator
product expansion, such decisive demonstrations are in fact lacking,
I believe.  Nevertheless, one cannot describe the passage from fluid
to glassy state fully in principle without invoking a diverging
number of degrees of freedom.  In such cases, more is necessarily
required beyond an accounting of local motional constraints.

\vspace{.7cm}

{\em
Q3) In spin glasses, the existence of a true spin glass phase transition has been well established by simulations and experiments. Do you believe that a similar result will ever be demonstrated for molecular glasses?
}

\vspace{.7cm}

It is not in our control whether the evidence will become as
convincing as it is for a spin glass.  It is in our control to devise
new realizations of glassiness and new ways to probe glassy systems
in the problematic region.  I think there is lots of scope for these
realizations and probes to give insight

\vspace{.7cm}

{\em
Q4) Why are there so many different theories of glasses? What kind of decisive
experiments you suggest to rule out at least some of them?
}

\vspace{.7cm}
To me the most plausible approach is to test the logical basis of
the theories eg mode coupling, with abstract models, chosen to
fulfill the hypotheses of the theories, but not to represent molecules

\vspace{.7cm}

{\em
Q5) Can you briefly explain, and justify, why you believe your pet theory fares better than others? What, deep inside, are you worried about, that could jeopardize your theoretical construction?
}

\vspace{.7cm}
My pet theory is the notion that the enabling motions that permit
fluidity are spatially extended vibrational modes with the properties
of weakly-localized wavefunctions. That is all I want for Christmas.
The justification is the form of the normal modes in a marginally
jammed granular pack, reported by Nagel et al and explained by
Matthieu Wyart.  Encouragement for this view comes from the extended
nature of the relaxation events in eg Dauchaut's grain pack
experiments.  The worry is that a satisfactory account of low-
temperature properties has to embrace two-level systems and
ultrasound echos.  These features seem to entail very local modes
with a finite set of energy states.

\vspace{.7cm}

{\em
Q6) In the hypothesis that RFOT forms a correct skeleton of the theory of glasses, what is missing in the theoretical construction that would convince the community?
}

\vspace{.7cm}
\textit{[Tom Witten did not wish to answer to this question]}

\vspace{.7cm}
{\em
Q7) (a) Exactly solvable mean-field glass models exhibit an extraordinary complexity
requiring impressive mathematical tools to solve them. Are you worried that for both the spin and structural glass problems it proves so hard to establish the
validity of mean-field concepts in finite dimensions? Why are finite dimensional versions of mean-field glasses always behaving in a very non mean-field manner?}

\vspace{.7cm}

Yes I am worried.  But is this so different from the development
of critical phenomena?  An advance in mathematical understanding was
needed, and at first the new things were not clearly grasped and so
they were hard to explain and distill into a simple form.

\vspace{.7cm}

{\em
Q8) In your view, do the recent ideas and experimental developments concerning jamming in
granular media and colloids contribute to our understanding of molecular
glasses, or are they essentially complementary?
}

\vspace{.7cm}

I think the jamming phenomena serve as a useful extension of
molecular glasses. Such extensions are surely needed in order to come
to a simple understanding of how mobility turns itself off.

\vspace{.7cm}

{\em
Q9) If a young physicist asked you whether he or she should work on the glass problem in the next few years,
would you encourage him or her and if so, which aspect
of the glass problem would you recommend him or her to tackle? If no, what problem in condensed matter theory should
he or she tackle instead? If yes, what
particular aspect of the glass problem seems the most exciting at present?
}

\vspace{.7cm}

Since the glass field is very mature, this young physicist should
be very smart and confident.  The essence of the problem is the
proliferation of local geometrical constraints.  I think there is a
lot of scope to devise artificial geometrical systems that embody is
idea and reveal how it works asymptotically.

\vspace{.7cm}

{\em
Q10) In twenty years from now, what concepts, ideas or results obtained on the glass transition in the last twenty years will be remembered?
}

\vspace{.7cm}

The simplest forms of mode-coupling theory, marginal jamming of
athermal systems, Quench memory phenomena

\vspace{.7cm}

{\em
Q11) If you met an omniscient God and were allowed one single question on glasses, what would it be?
}

\vspace{.7cm}
Are we on the right track at all (like early SU(3) leading to
modern quantum chromodynamics) or are we wandering in the wilderness
(like S matrix theory)?  (God would know what I meant)

\vspace{1cm}

\section{Peter G. Wolynes answers}

{\em
Q1) In your view, what are the most important aspects of the experimental data on the glass transition that any consistent theory explain? Is dynamical
heterogeneity one of these core aspects?
}

\vspace{.7cm}
At the very start, I think the most commonplace observation about structural molecular glasses requires explanation from a satisfying theory: the existence of sensibly rigid but amorphous objects! This rigidity is not an illusion but reflects a separation in timescales of individual molecular motions from collective motions--a separation spanning many orders of magnitude. Theories must explain how you get this rigidity without observing order, or show us how there is, in fact, broken spatial symmetry of some specific kind but how it is that we have so far failed to see it after very diligent efforts. An example of the first sort of explanation is the random first order transition theory based on the existence of truly aperiodic minima. In a sense the RFOT theory is the minimal generalization of the theory of ordinary first order transitions like crystallization to the situation in which there is a diverse set of structures each without periodicity \shortcite{1}.

The RFOT explanation for the initial origin of rigidity coincides with the explanation provided by the dynamic transition in mode coupling theory (MCT) \shortcite{2}. Both MCT and the density functional based form of RFOT theory are semiquantitatively accurate in predicting the plateau of the scattering function i.e., the Debye-Waller factor which quantitatively reflects the separation between vibrational and glassy motions. An example of the latter type of explanation based on a hidden broken symmetry that we generally don't observe is the icosahedratic theory of Nelson \shortcite{3} (Icosahedral order is seen in some metallic melts.) In my view postulating a priori models with activated events or defects in a rigid background from the get-go, fails to answer satisfactorily this primary question of the origin of rigidity. For this reason, I do not find merely postulating lattice models ``coarse grained'' in some manner at all intellectually satisfying on this most important conceptual point--how the coarse graining is done is the key to answering this fundamental puzzle.

The experimental investigation of structural glasses has revealed far more detailed facts that need explanation beyond the mere fact of rigidity, and since I believe in a unified theory I think it is important, ultimately, that no observation be left unexplained.

For example, I would require that an acceptable theory give a recipe, at least, for how to calculate any of its characteristic temperatures or energy scales from the intermolecular forces. This ``chemical'' attitude of mine is not universally shared in the physics community. I would not insist, however, on quantitative accuracy of such a calculational scheme at present since there are no small parameters apparent in the first formulation of the problem. Beyond this, a theory must explain three qualitative behaviors: non-Arrhenius temperature dependence, nonexponential time dependence and aging.

The non-Arrhenius temperature dependence of transport must be explained in a universal way. Those cases where Arrhenius behavior has been found can be continuously connected to non-Arrhenius systems by doping (e.g. adding a little $\mathrm{Na}_2\mathrm{O}$ to $\mathrm{SiO}_2$ changes silica from having seemingly Arrhenius behavior to having non-Arrhenius behavior.) Owing to this fact, I again think that merely postulating simple activated behavior at some elementary excitation level is a non-starter. Clearly separate theories for ``fragile'' and ``strong'' systems should be ruled out.

The nonexponential time dependent behavior of glassy molecular liquids must be explained, again in a universal way although special sources of heterogeneity may exist in some systems e.g. phase separation. Twenty-five years ago a hotly debated question was whether this nonexponential behavior implied heterogeneous patterns of motion in space or only in time. Experiments have finally answered this question quite clearly. This dynamical heterogeneity, is then, indeed, according to my taste, a core aspect of the glass phenomenon--in a sense it is the clearest manifestation of there being a true diversity of states. Elementary defect models are very hard pressed to explain the details of the spatial dynamical heterogeneity (they require kinetic constraints), even when such models can fit low order correlation functions.

A theory of glasses must give an explanation of the phenomena seen in aging systems, by definition, since the glass is a system no longer in equilibrium. The near-Arrhenius temperature dependence of aging in a nonequilibrium glass clearly points out the necessity of having activated transitions of some type emerging from the theory.

I would hold that, in addition, to the above absolute requirements for any theory of glasses, a satisfying theory should provide some answer to why the thermodynamic properties of supercooled liquids correlate as well as they do with their dynamical properties. This personal view again, is not held universally. Many theorists argue that the thermodynamic-kinetic connections noticed for decades, starting with Simon and Kauzmann more than half a century ago and emphasized for quite a while by Angell are just co-incidences or are the result of some ``law of corresponding states'' pointing to a common model of some unspecified sort. I do not see how practical minded people aware of the range of data already known can be satisfied with an explanation based on multiple co-incidences. Also why do the correspondences continue to pile up as more systems are studied? I would also put the burden on those advocating the mere corresponding states viewpoint to take a stance on the common model and show how to calculate numbers for a few systems. All evidence (e.g. pressure dependence, crosslink dependence of characteristic temperatures) points to the configurational entropy being the key quantity on which any law of corresponding states is based.

Along with this view of the importance of kinetic thermodynamic correlations, I believe a theory should tell us either how the Kauzmann entropy crisis (i.e. the impending vanishing of the configurational entropy at low temperature) is realized or precisely how it is avoided--without something happening, even the third law of thermodynamics comes into jeopardy with the total entropy for some systems extrapolating to zero only a bit below the conventional Kauzmann temperature.

\vspace{.7cm}

{\em
Q2) Why should we expect anything universal in the behaviour of glass-forming liquids? Is the glass transition problem well defined? Or is any glassy liquid glassy
in its own way?
}

\vspace{.7cm}

I was surprised when I first came across the fact that quantitative universal behavior was experimentally found for glass-forming liquids. One could imagine that after explaining the general origin of rigidity a satisfactory theory could find numerous classes of glassy behavior. For example if we took the icosahedratic picture as correct as the origin of rigidity in metallic glasses, we might imagine borosilicate glasses to find their rigidity to be based on some other platonic or  neo-platonic (?) form, van der Waals liquids another, etc. This multiplicity of mechanisms would be like the existence of the finite but large number of crystallographic group symmetries. Ordinary crystallization is quite sensitive to molecular details (e.g. the even-odd effect for alkane crystal melting points) but glass formation is much more robust and changes smoothly with composition. The glassy behavior of molecular liquids however seems to be quite universally expressed in terms of their configurational entropy $S_c$, and the fluctuations of configurational entropy as measured by their configurational heat capacity  $\Delta C_v$. These measures are sufficient to predict properties and otherwise one can be indifferent to the precise chemical and structural origin of the aperiodicity.

According to our understanding from the random first order transition theory, the observed quantitative universality for molecular liquids comes not from asymptotically close approach to a critical point but arises from the near universality of the Lindemann parameter measuring the scale of vibrational motions in any amorphous state. This Lindemann parameter universality is connected with the clear separation of vibrational motions from the slow motions in dense collections of particles and is predicted by density functional and microscopic mode coupling theories which are the starting points of the RFOT theory. It may be connected to geometrical features of space-filling structures in general. Still there are hints of an even stronger more universal behavior reflecting something like an approach to a real phase transition. The dependence of the complexity (i.e. the total configurational entropy) of a dynamically rearranging region on the time scale alone, for example follows from random first order transition theory in a general way without assuming a universal Lindemann parameter. Such a data collapse has been observed by Capaccioli et al. \shortcite{4}. Unfortunately it seems that those observations are not yet in a truly asymptotic regime of large correlation lengths, so a good but imperfect data collapse was found .I believe the observed mesoscopic scale of correlations confines the accuracy of such data collapse to about what we see in those experiments. A correlation range of five particle spacings as predicted by RFOT theory means accuracy in the 10\% range is about all that can be expected.

While molecular glasses have common patterns of behavior, different liquids do have individual characteristics because molecular liquids may manifest other phase transitions or near phase transitions in the deeply supercooled regime. The clearest examples are phase-separating glasses (like pyrex!!) but probably also include water, Silicon and pure silica ($\mathrm{SiO}_2$), for which there is strong circumstantial evidence of quite distinct amorphous forms. Some important substances, such as ``amorphous'' silicon may not be amorphous at all, but probably possess poorly developed periodic crystalline order (they are anisotropic often!). Despite the occurrence of these phase transitions deep in the supercooled state, I don't think merely saying everything is a result of an avoided phase transition quantitatively explains the universal patterns since one is clearly not in the asymptotic regime of any known avoided transition.

So, although I do not think each liquid will be glassy in its own way I believe molecular liquids will differ from each other quantitatively. Also if we go outside the class of structural glasses, more complex and distinct glassy behaviors will be found in systems such as colloids, granular assemblies and microemulsions. These need not be described by the same properties as the molecular glasses.

\vspace{.7cm}

{\em
Q3) In spin glasses, the existence of a true spin glass phase transition has been well established by simulations and experiments. Do you believe that a similar result will ever be demonstrated for molecular glasses?
}

\vspace{.7cm}

``Ever'' is a long time, so I hate to answer this question. Straightforward paths to a demonstration of a true phase transition for molecular glasses seem to be blocked by the near-universal relation between time scale and correlation length. Explicitly said, if we accept the predictions of random first order transition theory for molecular glasses or simply take the empirical results of Capaccioli et al. at face value it will be very difficult to achieve by cooling on human time scales an amorphous state with a large enough correlation length to be considered unambiguously in the asymptotic regime. The correlation length at a laboratory glass time scale of one hour reaches only five molecular diameters and waiting even years does not lengthen the scale much.

Are there routes other than cooling that could be used to prepare low configurational entropy or long correlation length glassy systems from molecules? Perhaps...., but nothing just on the horizon appears certain to go far. One promising laboratory route might be the preparation of glasses by vapor deposition. Ediger has prepared amorphous o-terphenyl with an apparent configurational entropy of about half that of a typically slow cooled glass \shortcite{5}.  Jake Stevenson and I have argued using RFOT theory that this is about all the further one can go in decreased configurational entropy \shortcite{6}. Indeed our arguments suggest the correlation length may not be much longer than in an ordinarily cooled glass. Would a correlation length of 10 satisfy people as being in the asymptotic limit? I don't know but I am sure more careful investigation of such hyperstable glasses merits the attention of theorists and experimenters alike.

As in the laboratory, computational paths to slower cooling rates are unpromising with classical computers. Moore's law is trumped by the Vogel-Fulcher law. True quantum computers may fare better but I suspect mere quantum simulator devices will also seriously suffer from glassy slowing effects despite some suggestions to the contrary in the literature \shortcite{7}.

A major mathematical advance in constructing computationally low energy aperiodic structures could also lead to a breakthrough on this score: rather than cooling maybe someone can find tricks for creating amorphous packings on the computer in a constructive way--a kind of super vapor deposition algorithm, a souped up version of Bennett's assembly algorithms. Another possible route on the computer, perhaps, could start by replicating finite examples of low energy structures, then doing some sort of detailed renormalization group-like construction, who knows? There are certainly practical motives arising from coding theory that suggest that exploring this option could lead to better error correcting codes; for this reason smart people are surely looking at this. The mathematicians did finally put to rest the centuries old Kepler conjecture on periodic fcc sphere packing in recent years, so optimism is not entirely silly.

Another approach would be to show the existence of true phase transitions in less direct ways. In contrast to the magnetic lattice models, series analysis tools have been underutilized for molecular fluids in my opinion. Does the Gaussian core model have a glass transition? Convincing use of such tools may be blocked by essential singularities connected with droplet excitations, however. Also searching for physical systems or mathematical models that interpolate between spin glasses and molecular glasses may help resolve the issue of whether there is a true phase transition.

I would like to take the liberty however to suggest that oversimplifying the question to one of the existence of a ``true phase transition'' misses the point of understanding glassy phenomena as we encounter them in the real world. Like flames, glasses are very often examples of ``intermediate asymptotics''. The ultimate state of most simple molecular liquids will be a periodic crystal a la the Kepler conjecture. We call configurations with few or small numbers of crystallites, glasses. Nevertheless in the thermodynamic, long time limit, strictly such crystalline configurations will dominate for simple liquids. We believe that a restriction to ``noncrystalline'' configurations somehow allows us to forget about this problem of delimiting the amorphous state but this may not be true. In any case suppose we could not partition off the crystalline part of the phase space sharply. Would the ``glass problem'' go away and be unworthy of study if the correlation length could only grow to 25 particle diameters before crystallization hit? I don't think so. Still it would be nice to know if there were such a spinodal limit for crystallization, I suppose. It might help a lot in materials synthesis.

\vspace{.7cm}

{\em
Q4) Why are there so many different theories of glasses? What kind of decisive
experiments you suggest to rule out at least some of them?
}

\vspace{.7cm}

Diversity and unity are the yin and yang of theoretical physics. We must always strive to achieve a balance. Multiple ways of looking at the same problem are essential both for present understanding and future theoretical progress--it is not better to look at the diffusion equation as describing a random walk than it is to view the equation as a continuum hydrodynamic theory, the two viewpoints illuminate each other. To reach balance, we must strive to see how diverse theories relate to each other, ultimately forming a united viewpoint. During the development of theories, imbalances between diversity and unity can often arise.

Creativity without discipline and laziness on the part of theoreticians are the main culprits that account for any unbalanced diversity of glass theories. Theorists are perhaps too good at coming up with theories and modifying them. That is a crucial part of our job, after all, but many theorists choose not to be disciplined either by conceptual simplicity or by experimental facts--if one believes that most experimental observations, for example the thermodynamic-kinetic correlations, are just coincidences because one is not studying the asymptotic regime, one is tempted to feel free to formulate theories that can fit data on a case by case basis using adjustable parameters. A very incomplete theory can seem plausible if one uses enough adjustable parameters. Also if the main conceptual problems are ignored by the starting point of the theory, for example by ignoring the amazing fact of amorphous rigidity by postulating rigidity from the beginning, numerous detailed models can be constructed to account, again on a case by case basis, for each specific piece of quantitative data.

Laziness presents two more serious problems. First, laziness makes it seem there are more theories than there really are. It is hard work to show how theories are related to each other, so seeking relations between theories is avoided by many people. Secondly I sometimes feel new elements are added to existing theories thus multiplying their diversity only to avoid work. I think professionals should deplore this kind of laziness more than we do. I feel you really should avoid throwing in a new exponent or adjustable constant if you do it just because you can't be bothered to calculate something. In this respect we have all been spoiled by the history of conventional critical phenomena. The remarkable scaling laws found in magnets, the liquid-gas transition, liquid helium etc. still tempt us to disdain the calculation of ``non-universal'' quantities--like the characteristic temperatures at which phenomena occur (that is supposed to be the business of engineers, I guess). But a claim of complete understanding really requires such calculations. In conventional critical phenomena the asymptotic regime was accessible and determination of the exponents seemed to be the most essential feature needed for understanding the basics--but for molecular glasses in the laboratory pre-asymptotic effects enter too. These pre-asymptotic effects are painful to calculate so we again try to avoid calculating them as long as we possibly can.

Computer simulations of ordinary phase transitions could be pushed into the regime of asymptopia more readily for conventional phase transitions than it has been possible to do for glassy ones. It is perhaps worth recalling however there was even confusion in the ``good old days'' of the early 1970's about whether simulations could give true phase transitions or not.

Not getting to asymptopia means we have a ``strong coupling'' problem. Other strong coupling problems, like the Mott transition in hard condensed matter physics and confinement in QCD, have suffered from unbalanced diversity and unity during their intellectual history too.

Vagueness of a theory when combined with laziness of theoreticians is a really big problem in ruling a theory out. If a model is vague it cannot be tested. Some of the diverse set of glass theories seem to me to be quite vague--that is another reason I recommend requiring their advocates to show a path back to the molecular forces! But also when the model is finally made definite you still must work out its consequences and live with them. I have explained my aesthetic reasons against being satisfied with facilitated Ising models with trivial thermodynamics--they assume a rigid background to start with, have many adjustable parameters, etc. But it is only recently that such models have been firmly ruled out by calculation: through hard work, Biroli, Bouchaud and Tarjus showed that these theories cannot explain the heat capacity of liquids and glassy kinetics with the same parameters \shortcite{8}. Nevertheless some of the salesmen of such theories just went ahead adding more adjustable parameters, ``renormalizing'' the difficulty away! This would be merely sad, except for the fact that so few people seem to have noticed what happened, judging from citations to the incorrect work.

You can see that I believe it is important to recognize that many specific theories being bandied about, have \emph{already} been falsified by experiment. These include pure mode coupling theory (without activated events) and facilitated lattice models with trivial thermodynamics. At the same time while shown to be incorrect or incomplete many elements of those theories as well as those of several other theories are appropriately left standing by experiment. Some elements of these theories are part of the RFOT construction, like the perturbative parts of mode coupling theory. Likewise, Nussinov has put forward arguments that connect the somewhat less well developed uniformly frustrated ``avoided'' phase transition models to random first order transitions \shortcite{9}. In this sense, distinguishing sharply between avoided transition models and RFOT theory is premature in my view.

As an example of another set of connections I note that rigidity percolation and even constraint counting a la Phillips also emerge when RFOT theory is used to treat network glasses \shortcite{10}. Thus again some parts of those theories are basically correct and consistent with my own thinking based on RFOT ideas.

Likewise the ``effects'' central to some models, such as ``facilitation''--the way in which a mobile region allows its neighbors to move more readily, are indeed real effects. A theory failing to account for these effects is incomplete. RFOT theory does give a way of describing facilitation effects as emergent phenomena. I have used that connection  recently in treating rejuvenation of glasses \shortcite{11}. Nevertheless I again emphasize that models that merely postulate facilitation from the start miss the main point of explaining the origin of rigidity.

Some attempts to treat activated events can be seen as variations on the mainline of thinking about RFOT theory. More than twenty years ago Randy Hall and I assumed a fixed size of activated event but treated the mismatch energy as growing as the system is compressed \shortcite{12}. I think this view doesn't explain the kinetic-thermodynamic correlation although it continues to re-surface as ``shoving'' models to use Jeppe Dyre's term \shortcite{13}. Ken Schweizer adds activated events to an RFOT-MCT framework but treats activated motions as largely being single particle events \shortcite{14}. I think this can only be correct near the crossover regime, and therefore is important for colloids but I don't think the approximation is good in the deeply supercooled regime.

Overall I believe the RFOT theory does well in balancing theoretical unity with diversity. I continue to spend my time trying to think up new experimental approaches that will illuminate the key features of molecular glasses. Most of these involve coming up with a more information rich experimental picture on the appropriate long time scales.

The biggest surprise from the RFOT theory is the size of the rearranging regions and how the CRR size scales with thermodynamics. While these aspects have been confirmed in a few cases by direct observation still more direct measurements of the size of the CRR's in a larger range of systems are needed to establish universality. Likewise understanding the shape of rearranging regions in supercooled liquids and glasses remains interesting. I believe directly visualizing amorphous systems on long time laboratory scales will eventually confirm the RFOT picture of relatively compact activated events in the deep glassy region that become more ramified near the crossover. Better imaging will rule out even more clearly point defect models than the thermodynamic argument has already done.
\vspace{.7cm}

{\em
Q5) Can you briefly explain, and justify, why you believe your pet theory fares better than others? What, deep inside, are you worried about, that could jeopardize your theoretical construction?
}

\vspace{.7cm}
Einstein is said to have remarked that it was curious that typically the only person who doubts the results of an experimentalist is the experimentalist himself and the only one who believes the results of a theorist is the theorist himself. In this spirit, I must say I have become quite convinced of the basic validity of the ideas in RFOT theory, despite being very much aware of gaps in its rigorous construction--it is made up of a set of approximations that work well for molecular liquids but that may break down for other systems. It has seemed to me that once you embark on constructing a theory of long lived aperiodic structures most of the key features of RFOT theory follow very naturally: A diversity of structures (i.e., configurational entropy) is almost a tautology. That configurational entropy must de-stabilize any single one of them seems self-evident to me and provides a natural driving force for activated events etc. Accepting these basic features still leaves many open questions but I think nearly uniquely sets you on developing the set of approximations made in RFOT theory as the ones of greatest simplicity. Other approaches such as frustrated phase transition scenarios, to the extent they differ from RFOT raise all sorts of questions of specificity that need answering before we can use them to illuminate our understanding of molecular liquids--please tell us what transition is being avoided! (Again the icosahedratic theory does this.) Likewise, I would ask excitation chain theorists: please tell us what you are being excited from, etc....

In any event, I think it is fair to say the RFOT theory is, at present, the most completely developed theory of structural glasses in a mathematical sense. The RFOT theory describes a path from the fundamental forces to experimental observables. The mode coupling theory also starts from the forces and has predicted interesting phenomena--the re-entrant glass transition of adhesive spheres etc. But, since the MCT dynamic transition is part of the RFOT framework this success from starting microscopically would be shared by an RFOT static  density functional approach. Most other theories of the glass transition don't yet say how to start from the intermolecular forces--leaving the problem of how to ``coarse grain'' liquid glass reality for future work.

The mathematical framework of the RFOT theory starts out in a very conservative fashion. Like the van der Waals theory of the liquid-gas transition, RFOT theory is acknowledged to be exact in the mean field limit of long range forces. Admittedly the mean field phenomena with replica symmetry breaking are richer than ordinary phase transitions but an extraordinary amount of beautiful work has illuminated this limit \shortcite{15}. This mean field theory provides a good basis for such concepts as effective temperature when systems are out of equilibrium.

In contrast to the liquid-gas transition where dynamics is a secondary feature, though, dynamics essentially connected to the finite range of the forces is important for the glass problem. The problem is to reconcile diversity with locality. Again, as anticipated by Kirkpatrick and myself more than twenty years ago, the Kac limit of long but not infinite range interactions can be analyzed rather completely \shortcite{16}. Silvio Franz has done beautiful work showing how the first corrections to the infinite range limit correspond to the simplest RFOT theory of activated events as instantons \shortcite{17}, paralleling but with much greater rigor recent work with Dzero and Schmalian \shortcite{18}.

The least developed part of the mathematical framework for the RFOT theory is the role of critical fluctuations that build on each other and thus go beyond what would be found perturbing around the infinite range. In the Kac limit, there are two correlation lengths, one of which diverges at the dynamic transition which without instantons can be made arbitrarily sharp by tuning the input range of the interactions. The other correlation length involves the instantons. This mosaic length scale diverges when the configurational entropy vanishes. In real molecular systems with short range forces these scales mix somehow, since the dynamical transition most definitely disappears as a strict divergence. Kirkpatrick, Thirumalai and I explored the consequences of assuming that there is only one divergent length \shortcite{19}. (The title of our paper was ``Scaling Concepts for the Dynamics of Viscous Liquids near \emph{an} Ideal Glass Transition''.) This single length hypothesis is perhaps an overly strong hypothesis but its specificity leads to simple predictions e.g. the Vogel-Fulcher scaling results as the only one consistent with a strict heat capacity discontinuity etc. There is evidence for multiple correlation lengths from the recent simulations of Biroli, et al. showing a thermodynamic signature of growing amorphous order \shortcite{20}. Also strictly speaking the explicit magnetic analogy proposed by KTW that was also explored recently by my research group allows multiple correlation lengths \shortcite{21}. The magnetic analogy suggests that an ideal glass phase transition occurs strictly near a point not a line in the phase diagram.

I think this question of the multiplicity of length scales is the origin of the mild controversy presently found within the RFOT-friendly community over the proper exponents to employ for the barrier scaling: the original ones of Kirkpatrick and mine which do not involve ``wetting'' or the ones that satisfy simple hyperscaling. The wetting free exponents are also those obtained from the literal single instanton calculation in the Kac limit.

Finally, but in my view, the most important reason I favor my ``pet theory'' is that the RFOT theory already works when it comes to the comparison with experiment! The strongest version of the theory (assuming only a single length and extrapolating mismatch energies to the molecular scale) makes numerous predictions about the correlation of thermodynamic data with kinetics that are well-satisfied. These correlations are predicted without the use of adjustable parameters. Many of these predicted correlations are discussed in detail in the review article I have written with Vas Lubchenko \shortcite{22}. They include in the classical regime:
\begin{itemize}
\item{the correlation of activation energy at the glass transition (``the fragility'') with the heat capacity discontinuity.}
\item{the prediction of the stretching exponent from the fragility}
\item{the prediction of how the ``non-linearity'' of the aging behavior in the glass correlates with fragility}
\item{the prediction of how the crossover temperature correlates with configurational entropy}
\end{itemize}

All of these predictions do not depend on there ever actually being a complete Kauzman entropy crisis--they depend on the local values of the configurational entropy and its derivatives. I am also surprised so many people object to using the Kauzmann temperature as at least a fiducial temperature since there is overwhelming evidence for at least a crossover at the Kauzmann point!

A very natural but admittedly bolder quantization of the RFOT framework made by Lubchenko and me also explains a large number of observations about the low temperature properties of glasses like two level systems and the Boson peak. Although those calculations are on much less secure footing than the purely classical arguments of the RFOT theory the confluence of those results with experiment also gives me confidence that the basic notions of the semi-classical RFOT theory are correct. I was very surprised that RFOT explained the weird universality of low temperature properties that had been emphasized by Tony Leggett. Indeed Lubchenko and I had been pursuing Leggett's interacting defect scenario with little success before we tried to use RFOT ideas. This history is at least a good example of why theorists believe their published theories when others do not--they have usually tried other alternatives and felt the pain of failure!

These numerous comparisons of the theory with experiment either are telling us the RFOT theory approach's approximations are reasonably sound for phenomena in the observed time and temperature range \emph{or} that there is some conspiracy of cancelling errors that is misleading us.

Except for such a possible conspiracy, then, I have no specific worries about what would jeopardize the RFOT theory construction in the observed experimental range for molecular liquids. I would share with many in the community however, worries about how RFOT theory will fare in ``asymptopia''--perhaps there is always some simple ordering that intervenes to prevent the ideal glass transition at low enough temperature. (How this would happen for an atactic polymer, I don't know!)--perhaps, the entropic droplet excitations themselves cut off the thermodynamic replica symmetry breaking transition, etc. as Michael Eastwood and I discussed \shortcite{23}etc.

Even with these worries, I would ask, is it then preferable to start the analysis of real data by an expansion about this ultimate asymptotic state rather than use measured thermodynamic quantities in the appropriate regime as input? In any event my own feeling is that any such approach would still lead to an RFOT-like theory as an approximation when the entropy density is not too low.

\vspace{.7cm}

{\em
Q6) In the hypothesis that RFOT forms a correct skeleton of the theory of glasses, what is missing in the theoretical construction that would convince the community?
}

\vspace{.7cm}

I don't know. The psychology of the community eludes me. Twenty-five years ago people told me they would be happy if there was a theory of structural glasses that would reproduce the macroscopic experimental trends, the more remarkable strange behaviors, and be exact in some limit. Those desires have been satisfied. Others, at that time, said they would like to see evidence for a growing length scale, which was predicted by RFOT theory in 1987. Evidence for the growing length has been directly found through nonlinear NMR experiments \shortcite{24} and through imaging approaches \shortcite{25}. Less direct determinations from rigorous inequalities related to macroscopic observables showing a growing correlation length agree even numerically with predictions of the strong form RFOT theory. Those predictions in the year 2000, were made years before the observations were analyzed \shortcite{26}. Still reviewers of papers using RFOT ideas demand another ``smoking gun'' from experiments.

I suspect some uneasiness about RFOT theory comes from the lack of comfort many people have with activated dynamics in general. With a handful of exceptions physical chemists really misunderstood transition state theory, even when they used it, until thirty years ago. Ordinary nucleation theory for ordinary first order transitions still confuses people, I have noticed. I believe all the results obtained using quasi-equilibrium Kramers theory like arguments in the existing RFOT theory can be obtained purely dynamically using Fokker-Planck equations or path integrals just as has been done for isomerizations in chemical physics. I don't know if achieving such comfort with reaction theory will be sufficient to convince the entire statistical physics community.

So you can see I am mostly at a loss as to what the community needs to be convinced. Perhaps the other interviewees can clue me in on what needs to be done and I can work on it!

\vspace{.7cm}
{\em
Q7) Exactly solvable mean-field glass models exhibit an extraordinary complexity
requiring impressive mathematical tools to solve them. Are you worried that for both the spin and structural glass problems it proves so hard to establish the
validity of mean-field concepts in finite dimensions? Why are finite dimensional versions of mean-field glasses always behaving in a very non mean-field manner?}

\vspace{.7cm}
Let me first make some remarks about your first statement that the exactly solvable mean-field glass models ``exhibit an extraordinary complexity requiring impressive mathematical tools to solve them''. This statement certainly characterizes how most people feel today. Yet at the very same time that I am impressed with the mathematical tools that have been invented to solve these problems, notably replica theory, I think our feeling of complexity about them is transitory. Any new formalism in theoretical physics seems complex to those who put it together and their contemporaries. Remember, again, it was Einstein who characterized one theory as ``a witches brew of determinants sufficiently complex so that it can never be falsified.'' Many would be tempted to say the same today about replica approaches to glassy phenomena. The theory Einstein was talking about was Heisenberg's quantum mechanics!

Replica methodology and its interpretation do indeed seem very intricate--to me too. But I can see younger workers have little trouble with them. I suspect with still more pedagogical effort coming generations will find the ideas and manipulations of replica theories like operator quantum mechanics to be quite natural. After getting that off my chest, let me answer your specific questions.

I am aware of the continuing problems of establishing which aspects of the replica symmetry breaking mean field theory apply to ordinary spin glasses in finite dimension. Although people seem agreed on the usefulness of mean field theory for estimating critical temperatures, at the next level of detail, everything seems to be discussed rather turbulently. Again for ordinary spin glasses everyone agrees something a bit beyond a simple phase transition enters both experiments and simulations: spin glasses exhibit significant aging, sensitivity to perturbations, etc. But there the consensus stops. Deciding whether these complex features reflect ``broken replica symmetry'' or ``chaos'' in a droplet picture or some hybrid of the two has taken much work. It has required developing precise definitions, doing big simulations, etc. I personally have not gotten too worked up about the controversy--primarily because I have not seen how the needed precision of argumentation helps us with the practicalities of understanding the problem of real random magnets or tells us how to use these ideas in other contexts like proteins, structural glasses, neural networks, etc. Perhaps I am a bit too American in my pragmatism, but it seems like the length scale at which the controversy rages keeps getting longer and longer with increasing years of study, making me feel that the controversy is more and more distant from being relevant to other problems where we might use the notions of glass theory.

Beyond this big controversy about replica symmetry breaking, still going back and forth between the Anglo-Saxons and the Continental Europeans focussed on ordinary Ising spin glasses, however, results more worrisome to me have appeared in the study of Potts glasses, whose lack of up-down symmetry, recommends them to us at the mean-field level as analogues of structural glasses. Several peculiarities seem to be at work conspiring to cloud the simulations of the Potts systems. First it was surprising how hard it was to even check the strict mean field theory out in the infinite range limit! Binder's calculations \shortcite{27} and simulations carried out by others show a particularly slow convergence to the thermodynamic limit with system size even for the infinite range Potts glass model. This convergence issue also seems to have surfaced in the recent work by Parisi on long-range models that mimic the Kac limit \shortcite{28}. Next even in the strict infinite range Potts glass sample-to-sample fluctuations seem to play a big role in the obvious estimators signalling the transition. This difficulty which occurs even for the infinite range limit must reflect itself even more starkly in tests of the RFOT mosaic scenario for finite range Potts glasses.

These worries coming from studying finite dimensional spin glasses do not, however, overwhelm my confidence in using RFOT ideas for structural molecular glasses for two technical reasons--both having to do with the ``hardness'' of the structural glass transition. Let me explain my terms.

Even in mean field random first order transitions are quite different from Ising spin glass transitions--the latter are ``soft'' while RFOT's are ``hard'' but to varying degrees depending on the system. My characterization of the Ising spin glass transitions as ``soft'' is my way of saying the Edwards-Anderson order parameters, q, varies continuously in magnitude at the transition (just as does the magnetization of a simple ferromagnet) at the transition value of q leads to a small restoring force for q fluctuations. This means the Ising spin glass is ``soft'' and yields all kinds of additional instabilities. Coping with these instabilities even at mean field put us on the long and winding road from the first mean field theories of Ising spin glasses through the radically brilliant one step replica symmetry breaking solution proposed by Bray and Moore for the Sherrington-Kirkpatrick model to the triumphant Parisi hierarchical ansatz! At finite dimension the same sort of instabilities (tamed globally by the Parisi ansatz) lead to a zoo of critical replicon modes according to de Dominicis and to the annoyingly high upper critical dimension $d=6$ or $d=8$ according to Fisher. These soft modes are also at the core of the difficulty of testing the mean field constructs in less than 8 dimensions for the Ising system.

The situation should be better for Potts glasses since their mean field transition is ``hard''--the Edwards-Anderson order parameter is discontinuous. This means no unstable modes appear at the mean field transition--Bray and Moore's single step of RSB, inadequate for Ising systems, is enough--and everything should be wonderful! But, alas, there are droplets!!

The problem is that the discontinuity of $q$ for Potts glasses while strictly present is small when the number of components is not huge! The transition is, while technically ``hard'', a bit squishy. The small discontinuity in q does formal miracles for those who love to use Landau expansions but it means the transition is effectively nearly soft: the interface energy cost for making a droplet is very small so entropic droplets proliferate and interact with each other. Michael Eastwood and I made estimates suggesting these droplet excitations should have a big influence on the configurational entropy made for Potts glasses --enough perhaps to endanger the thermodynamic random first order transition \shortcite{23}. This view would also be consistent with what we find by constructing the explicit magnetic analogy a la Stevenson et al. \shortcite{21}. In the magnetic analogy, lowering the average surface tension as happens for a not very hard transition, while keeping its fluctuations and the field fluctuations (analogous to configurational entropy fluctuations) fixed wipes out the replica symmetry breaking transition. When the transition is wiped out an analysis along the lines of Tarzia and Moore \shortcite{29} would seem to be a good starting point.

Because of the small Lindemann parameter, structural glasses are much ``harder'' than are the Potts glasses with modest numbers of components that have been studied. The analogue of the Edwards-Anderson order parameter, the plateau, in the first peak of the Structure factor as revealed by neutron scattering is nearly about .9. This means the transition is quite hard, moving the surface costs of entropic droplets way up, so they cannot significantly renormalize the impending Kauzmann entropy crisis. The large activation free energy of entropic droplets, however, slows motions down as it must, making simulations of structural glasses difficult to do in the deep glassy regime.

I think the unusual ``hardness'' of molecular structural glasses is at the root of the quantitative success of the simplest RFOT theory as it currently stands. The existing RFOT theory just adds a non-perturbative decoration of droplets to the mean field theory. This level of calculation is OK for structural molecular glasses but many other ``glassy'' systems will not be so easily analyzed since they are softer. We must learn to cope with all the issues of critical fluctuations, replica Goldstone modes, etc. etc. if we want to completely understand such systems as glassy microemulsions, foams, etc.

\vspace{.7cm}

{\em
Q8) In your view, do the recent ideas and experimental developments concerning jamming in
granular media and colloids contribute to our understanding of molecular
glasses, or are they essentially complementary?
}

\vspace{.7cm}

The study of jamming is important in its own right. The experimental results showing very strong divergences near jamming also seem crisp and therefore are especially intriguing. None of them made sense to me in the context of molecular glasses until the recent work of Mari, Krzakala and Kurchan \shortcite{30}, which was a real revelation. The connection they found between jamming with an extreme sort of nonequilibrium aging in which the pressure becomes infinite in the mean field limit opens up the connection to the RFOT viewpoint on molecular glasses.

I think the replica approach makes clear, however, that the origin of the rigidity of near equilibrium thermal glasses is not precisely the same as that for far from equilibrium jamming systems however. The single J-point formalism which has been discussed so widely therefore seems way too oversimplified.  The replica theory has a continuum of jammed structures. Much more work needs to be done however to go beyond mean field theory in uniting jamming rigidity which reflects far from equilibrium aging with the rigidity of glasses which is a near to equilibrium. Clearly additional theoretical innovations will be needed since the analogue of fictive temperature is not set up so simply for a jammed granular or colloidal system as it is in a cooling protocol of a molecular liquid. Probably there is some sort of turbulence cascade like viewpoint (a la Kolmogorov) that will tell us how fictive temperature is generated and transported in these driven systems from large scales to short or vice versa. Perhaps more radical ideas involving intermittency or avalanches are needed. Progress on understanding driven jammed glasses will be essential to understanding structure formation in biology at the supramolecular level--the inside of cells is constantly being built and rebuilt alongside the usual diffusive motions of the individual macromolecular inanimate colloids which make up the cytoplasm. I consider exploring the connection between molecular glasses and jammed colloidal systems therefore a high priority.

\vspace{.7cm}

{\em
Q9) If a young physicist asked you whether he or she should work on the glass problem in the next few years,
would you encourage him or her and if so, which aspect
of the glass problem would you recommend him or her to tackle? If no, what problem in condensed matter theory should
he or she tackle instead? If yes, what
particular aspect of the glass problem seems the most exciting at present?
}

\vspace{.7cm}
Yes, absolutely. In my view coping with diversity is the key problem of glassiness and is the key issue in many areas of theoretical physics. The ``single state'' viewpoint still dominates throughout theoretical physics. I view this single state tyranny as a road-block in many areas of physical, biological, and social science. We need young people to join the revolution. Examples where better coping with diversity would help include:

\begin{itemize}
\item{``Hard'' condensed matter physics. The paradigms of Sommerfeld, Peierls and Landau still dominate--nearly everything is an attempted perturbation about the single state Fermi liquid or some single ordered structure. Those a bit more daring add two states or a quantum critical point to their thinking. Joerg Schmalian impressed me many years ago that we must take seriously the glassy behavior seen in the phase diagram of cuprates \shortcite{31}. Dobrosavljevic has gone further in suggesting a glassy system might provide the infamous ``quantum critical point'' so-desired by many as an explanation of the pseudo-gap phase \shortcite{32}.

    When I recall the ancient controversy in chemistry between whether molecules exhibit tautomerism (where different structures isomerize classically) or exhibit mesomerism (where different structures quantum mechanically superimpose via resonance), I come to feel that all the resonating valence bond work starting with Anderson, might make more sense when viewed via the energy landscape language of glassy systems. }
\item{Particle physics and cosmology--Landscapes have entered these subjects leading to much controversy, even in the popular press \shortcite{33}. When I read research papers in these areas I feel that this community is still groping with how to deal with diversity, and could learn something from the efforts of the condensed matter community in studying glasses. In any case I am grateful to David Gross and his collaborators, already, for their early work on mean field spin glasses which helped RFOT theory get started. Why such a distinguished particle physicist worried about such a question is unclear. I will have to ask him!}
\item{Biological Physics--The study of protein dynamics and particularly protein folding continues to be influenced by glass physics. I have found this to be a two-way street for information flow.

    But diversity is a key issue in all of biological physics not just at the single molecule level of proteins. Problems where diversity matters range from gene regulation (Are cell types minima on a landscape?~\shortcite{34}), the structure and dynamics of individual cells (what determines the shape of a cell?) and, of course, the problem of memory where already a vigorous interchange with glass physics was developed a quarter century ago through the pioneering work of Hopfield.}
\item{Econophysics: Didn't Keynes make it clear there were multiple states in the economy? His statement that ``In the long run, we are all dead!'' is a classic appreciation of the importance of long lived i.e. glassy states. Multiple solutions to satisfying individuals in their social and economic interchanges are obviously possible. Bouchaud's writings on this type of issue are very inspiring \shortcite{35}.}
\end{itemize}

In all of the above areas of theoretical science the battle even to acknowledge diversity as a factor has just begun. People really want to fit systems into the Procrustean beds of the single state techniques they learned in school years ago. At the same time they feel no progress in understanding diversity has been made in any problem which clearly exhibits diversity. Thus one of the psychological hold-ups to progress is really settling the issue of what's going on in molecular glasses. I think getting complete control of the molecular glass problem is therefore widely important to the progress of physical theory as a whole.

I believe a young physicist who wants to work on any challenging problem in physics will eventually have to learn about glasses. She or he may also be able to directly contribute to the study of glasses and supercooled liquids, proper, while doing so.

In the study of molecular glasses over the next years four areas of enquiry stand out in my mind. The first is how to exploit the existing universal ideas from the RFOT theory to address the chemical and molecular details of liquids and amorphous materials: Water, that most biologically relevant and reactive solvent, has already been explored keeping in mind its amorphous state(s). I think the ideas from RFOT theory can help clarify many problems here. Amorphous semiconductors have exciting electronic properties for which local bonding ``defects'' are crucial How can we look at such defects in a local energy landscape picture? Vas Lubchenko has exciting ideas in this question.

Second, the properties of aging and rejuvenating systems need much more study both in the laboratory and from a formal theoretical point of view. Is time reparametrization symmetry an actual but emergent property away from the mean field limit?

Third, the truly intermediate regime around the crossover temperature may have a deeper representation than in the current RFOT mean field droplet approach. Within RFOT theory this transition comes about from the ramification of the entropic droplets and resembles the Hagedorn transition of string theory \shortcite{36}. Is there a way to see strings formally as real dynamic entities with microscopically calculated interactions? I am not satisfied with all the approximations made in Langer's theory. Is there a way to picture some sort of duality between the situation above and below the crossover--mobile strings below morphing into partially rigid structures above?

Fourth, the origin of amorphousness still needs clarification. RFOT theory takes aperiodic minimal energy structures as given. Simulations show aperiodic minimia exist so that is enough. Are these minima just superpositions of defects a la the icosahedratic theory for simple fluids? Or is such a view something like Landau's construction of turbulent states via multiple periodic instabilities. He thought a system becoming turbulent just continued bifurcating but instead strange attractors were found at the onset of turbulence. Like turbulence, is spatial chaos intrinsic and more subtle?

There is plenty to do and time is a-wastin'!

\vspace{.7cm}

{\em
Q10) In twenty years from now, what concepts, ideas or results obtained on the glass transition in the last twenty years will be remembered?
}

\vspace{.7cm}
Results from the last 20 years or so should already be part of the curriculum. The mere idea of an existence of an energy landscape with great diversity is noncontroversial and is a powerful way even for undergraduates to picture many systems ranging from liquids to proteins. There is no reason not to teach about some formal aspects of glass transition theory: strict mode coupling theory (as a re-summed perturbation theory), density functional treatments of aperiodic minima, basic replica methods and how the entropy crisis occurs in simple models. No one really disputes these are mathematically correct any more and these ideas are the starting point for future advances. I think the entropic droplet and mosaic notions may be modified but will survive further examination. The future of current issues including the critical phenomena aspects and effects of interactions between activated events (``facilitation'') will I hope be soon enough settled to be remembered. Simple facilitated models will survive, not because of their direct applicability to molecular liquids, but because of their pedagogic simplicity and applicability to other systems.

\vspace{.7cm}

{\em
Q11) If you met an omniscient God and were allowed one single question on glasses, what would it be?
}
\vspace{.7cm}

The \textit{New York Times} has already quoted one scientist as derisively referring to RFOT theory as ``metaphysical'', so I really don't want to go further by entering into theology. My own thinking parallels a joke that made the rounds in America a few years ago after the truly tragic mismanagement of the Hurricane Katrina relief:

A devout believer was living in the predicted path of the Hurricane. TV reports urged all those in the path to evacuate to safer ground. The believer did not go saying ``I trust in the Lord. He will save me.'' The flooding began and the water came up to the level of the porch. At this time the sheriff came by in a boat, asking the believer to come with him to safety. The believer did not go, saying, ``I trust in the Lord. He will save me.'' The water continued to rise, finally reaching the roof of the house. The believer had already climbed out onto the roof. A helicopter appeared and the National Guard lowered a ladder telling the believer to come on up to be whisked away. The believer stayed on his roof citing his trust that the Lord would save him.

The believer drowned.

Being a true believer he did go to Heaven. There he asked God, ``Why didn't you save me?'' The Lord answered ``I allowed meteorologists to predict the storm's path. You didn't listen. I sent the sheriff with his boat to get you. You didn't listen. Finally I sent a helicopter. I really don't think it is fair to ask me that question!''
\vspace{1cm}

\subsection{Reflections on the interview responses, by Peter G. Wolynes}

Despite differences of tone and emphasis between the respondents I think there are many aspects of the glass transition on which there is agreement. Many of the issues raised by Langer, Witten and Kurchan were already discussed by me in my first response. I limit myself primarily here to directing the reader to specific papers that discuss a few more of the issues raised by the other responses.

Langer spells out some specifics about his discomforts with RFOT theory. Many of his worries have already been mentioned in existing papers on RFOT theory. Doubtless these issues haven't been completely resolved to everyone's satisfaction but should provide an agenda for future work or reformulations.

It is worth pointing out that the early RFOT theory papers contrasted the nucleation like arguments of RFOT theory and those for ordinary first order transitions and emphasized that there were differences as to growth of unstable phases etc. The reformulation of the nucleation style arguments by Lubchenko and Wolynes was aimed at aging \shortcite{10000} but is indeed quite consistent with the contemporary Biroli-Bouchaud arguments for equilibrated samples and makes clear why the problem is distinct from the traditional nucleation of an unstable phase. The Lubchenko-Wolynes paper discusses in a critical way the quasi-equilibrium assumption of the RFOT theory:  RFOT theory does indeed postulate that the only constraints on motion across the transition state for escape from a minimum are those that are in fact generated by the direct interactions. Quantitatively these constraints are accounted for in the density functional starting point. Indeed other additional topological constraints may enter for some systems--entangled polymers were noted as an example by LW. ``Short cuts'' to escape from a present local minimum would also not satisfy the quasi-equilibrium assumption: such short cuts were discussed by the LW paper as well in the language of defect models. Specifically the role of string-like excitations that can give short cuts for minimum escape has been discussed by Stevenson, Schmalian and myself \shortcite{20000} and their importance to secondary relaxation has been explored by us recently \shortcite{30000}.

Witten raises the issues of soft modes and two level systems which are apparent at cryogenic temperatures. Although my remarks were aimed at the high temperature, nearly classical phenomena, the RFOT theory does make some predictions about the relevant quantum low temperature properties which depend on dynamic modes see Lubchenko and Wolynes, Phys. Rev. Lett. \shortcite{40000}, PNAS \shortcite{50000} and a long article in Advances in Chemical Physics \shortcite{60000}.

I found Kurchan's responses very much on the mark and agree with the gestalt of his description of the RFOT approach. I particularly enjoyed his metaphor of the RFOT theory as a building with the ``upper floor made of wood for lack of budget.'' (May I quote this next time I apply for research funds?) In this regard, however, I would remind seekers of the Hold Grail of glass physics of the lesson taught by Henry Jones, Jr. in the film ``Indiana Jones and the Last
Crusade'': the power of the Grail is not entirely determined by its materials of construction.  Choose wisely!



\bibliographystyle{OUPnamed_notitle}
\bibliography{refs_chap1_corrected}
\end{document}